# Evidence for Exciton Crystals in a 2D Semiconductor Heterotrilayer

Yusong Bai,[1,#,†] Yiliu Li,[1,#] Song Liu,[2] Yinjie Guo,[3] Jordan Pack,[3] Jue Wang,[1] Cory R. Dean,[3] James Hone,[2] X.-Y. Zhu[1,*]

1. Department of Chemistry, Columbia University, New York, NY 10027, USA.
2. Department Mechanical Engineering, Columbia University, New York, NY 10027, USA.
3. Department of Physics and Astronomy, Columbia University, New York, NY, 10027, USA.

**Two-dimensional (2D) transition metal dichalcogenides (TMDC) and their moiré interfaces have been demonstrated for correlated electron states, including Mott insulators and electron/hole crystals commensurate with moiré superlattices[1–7]. Here we present spectroscopic evidences for ordered bosons - interlayer exciton crystals in a $WSe_2/MoSe_2/WSe_2$ trilayer, where the enhanced Coulomb interactions over those in heterobilayers have been predicted to result in exciton ordering[8–10]. While the dipolar interlayer excitons in the heterobilayer may be ordered by the periodic moiré traps[11–14], their mutual repulsion results in de-trapping at exciton density $n_{ex} \geq 10^{11}$ cm$^{-2}$ to form mobile exciton gases and further to electron-hole plasmas, both accompanied by broadening in photoluminescence (PL) peaks and large increases in mobility[15]. In contrast, ordered interlayer excitons in the trilayer are characterized by negligible mobility and by sharper PL peaks persisting to $n_{ex} \sim 10^{12}$ cm$^{-2}$. We present evidences for the predicted quadrupolar exciton crystal and its transitions to dipolar excitons either with increasing $n_{ex}$ or by an applied electric field. These ordered interlayer excitons may serve as models for the exploration of quantum phase transitions and quantum coherent phenomena.**

Fermions and bosons can undergo quantum phase transitions and form ordered ground states. These processes are determined by changes to fundamental parameters in the system Hamiltonian[16] rather than temperature in the more familiar thermal phase transitions. TMDC homo- and hetero-bilayers have emerged as model systems for the realization of ordered electron

---

[*] To whom correspondence should be addressed: xyzhu@columbia.edu
[#] These authors contributed equally.
[†] Present Address: Department of Chemistry, Brown University, Providence, RI 02912, USA

states. The reduced screening can give rise to strong many-body Coulomb potential dominating over the carrier kinetic energy, which is intrinsically low in these systems and is further lowered by moiré band flattening, leading to ordered fermions[1–7]. In contrast, experimental realization of the ordering of the bosonic excitons in 2D has remained a challenge[17]. Interlayer excitons in TMDC heterojunctions with type-II band alignment may offer a platform for realizing ordered excitons. The spatial separation of the electron-hole pair results in long exciton lifetimes ($10^2$ ns) which in turn permits broad tuning of exciton density under continuous-wave (CW) optical excitation[18]. The large out-of-plane dipole moments enable long-range Coulomb interaction, a key ingredient in Wigner crystal formation[19] from their Fermionic counterparts. The presence of moiré pattern can also flatten the electron/hole bands[20,21] and introduce potential traps to localize interlayer excitons[22], thus reducing exciton kinetic energy and facilitating crystallization. Despite these favorable conditions, interlayer exciton crystals in TMDC heterobilayers have not been observed to date.

Interestingly, recent theoretical studies predicted that increasing the thickness from an asymmetric heterobilayer to a symmetric heterotrilayer (e.g., $WSe_2/MoSe_2/WSe_2$) results in the formation of robust crystalline phases of exciton[8–10]. Fig. 1a illustrates the calculated band structures[8] for $MoSe_2/WSe_2$ and $WSe_2/MoSe_2/WSe_2$. In the latter, there is an additional valence band splitting ($\Delta_\pm$) in the K valley due to hole-hoping between the two $WSe_2$ layers. As a result, the interlayer exciton is quadrupolar ($IX^{\Delta_Q}$) and can be considered as consisting of 0.5 hole each in the top and bottom $WSe_2$ layer and one electron in the middle $MoSe_2$ layer. A quadrupolar exciton is stabilized over a dipolar exciton ($IX^{u(d)}$) by $\sim$ -20±10 meV, which is termed the "quantum fluctuation" energy[8]. Following Slobodkin et al.[8], we write the many-body Hamiltonian of interlayer excitons in the $WSe_2/MoSe_2$ heterobilayer as:

$$\widehat{H}_{BL} = \sum_i (K_i) + \sum_{i<j} V_{\uparrow_i \uparrow_j} \qquad (1),$$

where $K_i$ the exciton kinetic energy operator, $V_{\uparrow_i \uparrow_j}$ is the Coulomb repulsion between parallelly ($\uparrow_i \uparrow_j$) aligned dipoles. The short-range exchange interaction due to Fermionic characters of the charge separated excitons is neglected at low densities. For the symmetric $WSe_2/MoSe_2/WSe_2$ heterotrilayer, the Hamiltonian becomes:

$$\widehat{H}_{TL} = \sum_i (K_i) + \sum_{i<j} \left( V_{\uparrow_i \uparrow_j} - V_{\uparrow_i \downarrow_j} \right) - \Delta_Q \sum_i \sigma_i^x \qquad (2).$$

where the additional $-V_{\uparrow_i \downarrow_j}$ accounts for Coulomb attraction when two interlayer excitons anti-parallelly ($\uparrow_i \downarrow_j$) aligned in the top and bottom heterobilayer, respectively; the last term is the quadrupolar exciton stabilization energy and the $\sigma_i^x$ operator flips each dipole.

The additional two stabilization terms in the heterotrilayer are essential to the formation of robust interlayer exciton crystals, shown in the phase diagram in Fig. 1b. Here, the competition between $\Delta_Q$ and the Coulomb energy $U \propto \frac{e^2}{\epsilon d}$ (*e*: electron charge, *ε*: dielectric constant, *d*: interlayer spacing) is represented by a unitless parameter $R = \frac{\Delta_Q}{U}$. For the specific $WSe_2/MoSe_2/WSe_2$ trilayer system[8], R is estimated to be above $R_c$, and quadrupolar excitons adopting a triangular lattice is the lowest energy state at exciton density $n_{ex} < 10^{13}$ cm$^{-2}$. Increasing $n_{ex}$ results in a phase transition to the staggered dipolar excitons in a square lattice where Coulomb attraction dominates. For low *R* values, staggered dipolar exciton droplets may also form at low $n_{ex}$. Note that the predicted exciton lattices illustrated in Fig. 1b differ from the recently reported exciton insulators[23,24] in a TMDC moiré bilayer separated from a TMDC monolayer or exciton density waves[25] in a double TMDC moiré bilayer with weak inter-moiré layer coupling. In these systems, the correlation effect is likely rooted in the Mott physics of the electron/hole residing in the moiré structure.

Here we report spectroscopic evidences for the predicted exciton crystals and quantum phase transitions in the $WSe_2/MoSe_2/WSe_2$ trilayer. We fabricate two nominally identical devices, each consisting of a $WSe_2/MoSe_2/WSe_2$ heterostructure with an extra $WSe_2/MoSe_2$ bilayer region as control using the highest quality $WSe_2$ and $MoSe_2$ monolayers[26] and the transfer stacking technique detailed elsewhere[1] (see Methods Figs. S1-S6). We present results mostly from device-I in the text and those from device-II in Supporting Information. All main findings are qualitatively reproduced in the two devices, apart from variations expected from sample-to-sample heterogeneity. A schematic of the structure is shown in Fig. 1c and an optical image in Fig. 1d. The three TMDC monolayers are at near-zero twist angles (<1°) as characterized by phase-resolved second-harmonic-generation (SHG). At such a small twist angle, the moiré unit cell density is $n_{moiré} \leq 3.2 \times 10^{11}$ cm$^{-2}$. We adopt a dual-gate architecture with hexagonal boron nitride (h-BN) encapsulation and semitransparent graphene (3-4 layers) top ($V_{tg}$) and bottom ($V_{bg}$) gate electrodes; this allows us to independently control the out-of-plane electric field and electrostatic doping of the heterostructure without losing optical access (Methods). We probe the exciton phases in the trilayer via PL spectroscopy. The near-perfect angular alignments satisfy electron-hole momentum-matching conditions for optical transitions[27].

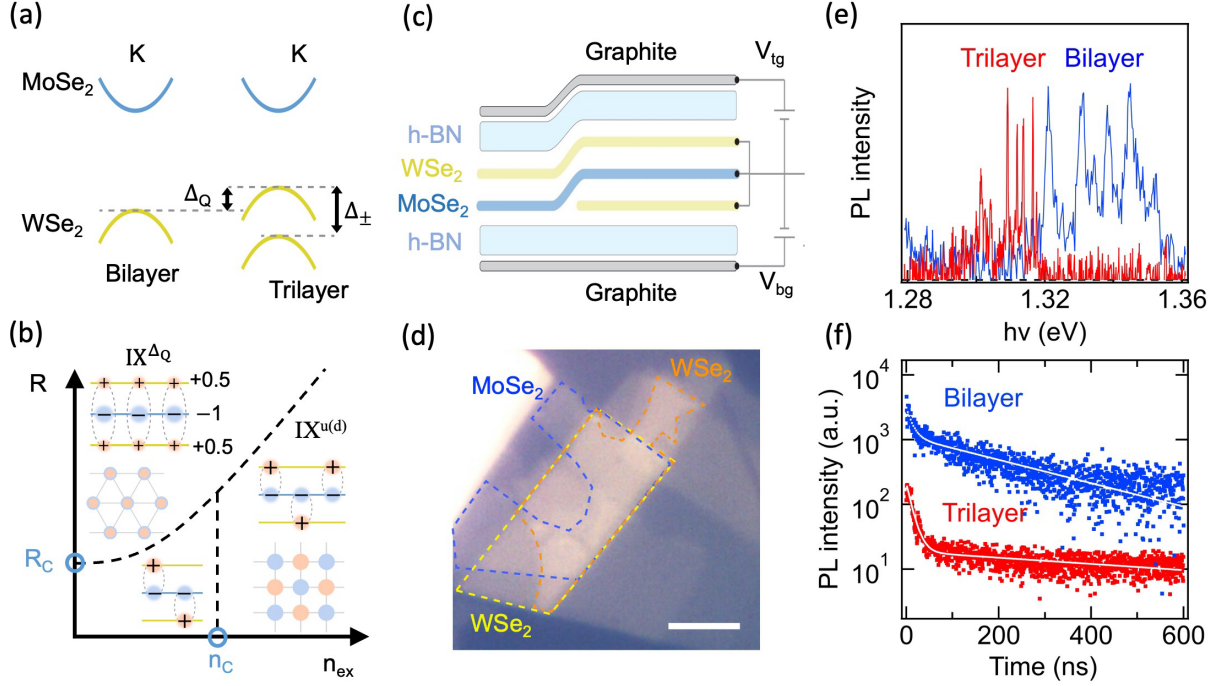

**Fig. 1 | The WSe$_2$/MoSe$_2$/WSe$_2$ trilayer: energetics, excitonic phases, and device architectures**. **(a)** Band structure of the WSe$_2$/MoSe$_2$ bilayer and the WSe$_2$/MoSe$_2$/WSe$_2$ trilayer. $\Delta_\pm$ is the hole-hopping-induced band splitting and $\Delta_Q = \frac{1}{2}\Delta_\pm$ is the energy difference between the valence band edges of the bilayer and the trilayer. **(b)** Schematic phase diagram of interlayer excitons in WSe$_2$/MoSe$_2$/WSe$_2$. $n_{ex}$ is the exciton density and R ($= \frac{\Delta_Q}{U}$; $U$, the Coulomb interaction) is a dimensionless parameter. IX$^{\Delta_Q}$: the quadrupolar interlayer exciton crystal; IX$^{u(d)}$: the anti-parallel interlayer excitons with dipole moments pointing upward (u) or downward (d). **(c)** Schematic of the device. $V_{tg}$ and $V_{bg}$ are the top- and bottom-gate voltages for independent control over electric field and electrostatic doping. **(d)** An optical image of the WSe$_2$/MoSe$_2$/WSe$_2$ heterostructure (device-I) in top view. Scale bar is 10 $\mu$m. A larger image is in Fig. S5. **(e)** PL spectra of interlayer excitons from device-I in the WSe$_2$/MoSe$_2$ bilayer (red) and the WSe$_2$/MoSe$_2$/WSe$_2$ trilayer (blue) at $V_{tg} = V_{bg} = 0$ V; CW excitation, h$\nu$ = 2.33 eV; calibrated exciton density $n_{ex}$ = 2.5x10$^{10}$ cm$^{-2}$ and 3.7x10$^{10}$ cm$^{-2}$ for the bilayer and trilayer, respectively; (f) Time resolved PL from interlayer excitons in device-II in the WSe$_2$/MoSe$_2$ bilayer (blue dots) and the WSe$_2$/MoSe$_2$/WSe$_2$ trilayer (red dots) regions at $V_{tg} = V_{bg} = 0$ V; pulsed excitation (h$\nu$ = 1.65 eV, pulse width ~ 150 fs); $n_{ex}$ ~ 8x10$^{12}$ cm$^{-2}$ for both regions. White curves are bi-exponential fits with time constants $\tau_1$ = 12±2 ns and $\tau_2$ = 230±30 ns for the bilayer and $\tau_1$ = 12±2 ns and $\tau_2$ = 900±100 ns for the trilayer. Sample temperature T = 4.6 K in all measurements.

Fig. 1e compares PL spectra at zero electric field and gate-doping ($V_{tg} = V_{tg} = 0$) and low excitation densities, $n_{ex} = 2.5 \times 10^{10}$ and $3.7 \times 10^{10}$ cm$^{-2}$, for the bilayer and trilayer, respectively. Calibration of exciton densities can be found in Supporting Information and Fig. S7-S9. Each spectrum in Fig. 1e consists of a distribution of narrow peaks in the interlayer exciton energy range. In the bilayer region, the sharp PL peaks with full-width-at-half-maximum (FWHM) of 1.0±0.2 meV result from interlayer excitons in the quantum-emitter like moiré traps that preserve $C_{3v}$ local symmetry[11,13,28] and exhibit photon antibunching characteristics[29]. Variations in peak energy may be attributed to dielectric and/or strain disorder. In the trilayer, the PL peaks are five-times sharper with FWHM = 0.20±0.05 meV. The sharp PL peaks in the trilayer persists to $n_{ex}$ much higher than they do in the bilayer, as detailed below. Note that the sharp emission peaks from TMDC heterostructures may also come from extrinsic localization potentials[30,31]. Interlayer exciton emission from these extrinsic traps should saturate when $n_{ex}$ exceeds the trap density, which is expected to be $\ll n_{moiré}$ in our samples[13,26]. As detailed below, the sharp PL peaks persist to $n_{ex} > n_{moiré}$ and remain identifiable at $n_{ex} \geq 10^{13}$ cm$^{-2}$ despite broadening. Thus, we attribute the sharp PL peaks as originating from moiré exciton traps[11,13,28,29].

The transition dipole moment of dipolar interlayer exciton in each WSe$_2$/MoSe$_2$ heterobilayer is dominantly (99±1%) in-plane[32]. For the predicted quadrupolar interlayer exciton in the WSe$_2$/MoSe$_2$/WSe$_2$ heterotrilayer, we can consider an out-of-phase and an in-phase combination of two in-plane transition dipoles. The former leads to net zero transition dipole moment; only the latter is optically bright. Thus, we expect a reduction in total oscillator strength going from dipolar to quadrupolar interlayer excitons. This is confirmed in time-resolved PL measurements, Fig. 1f, for an initial excitation density of $n_{ex} \sim 8 \times 10^{12}$ cm$^{-2}$ for both the bilayer (blue dots) and the trilayer (red dots), with biexponential fits shown as white curves. In the bilayer, there is an initial fast PL decay characterized by a time constant $\tau_1 = 12\pm2$ ns followed by a slower time constants of $\tau_2 = 230\pm30$ ns (see Fig. S10 for more starting $n_{ex}$ values). The former is attributed to many-body scattering which accelerates interlayer exciton recombination at high densities[18] and the latter is in quantitative agreement with prior measurements of the WSe$_2$/MoSe$_2$ heterobilayer[18,33]. In the trilayer region, the initial decay is again characterized by $\tau_1 = 12\pm2$ ns while the majority of the exciton population (>85%) decays at $\tau_2 = 900\pm100$ ns. The latter is 4x lower than that in the bilayer.

Note that earlier work by Choi et al. on the WSe$_2$/MoSe$_2$/WSe$_2$ system reported higher PL intensity of interlayer excitons from the trilayer than that from the bilayer region[34]; this result was cited in the work of Slobodkin et al.[8] However, the interpretations of Choi et al. were likely incorrect because the samples used were of extensive disorder and poor van der Waals contacts, as evidenced by the broad PL peaks and the presence of emission from intralayer exciton[34]. Intralayer exciton emission in TMD heterojunctions should be completely quenched[11,13,18,33].

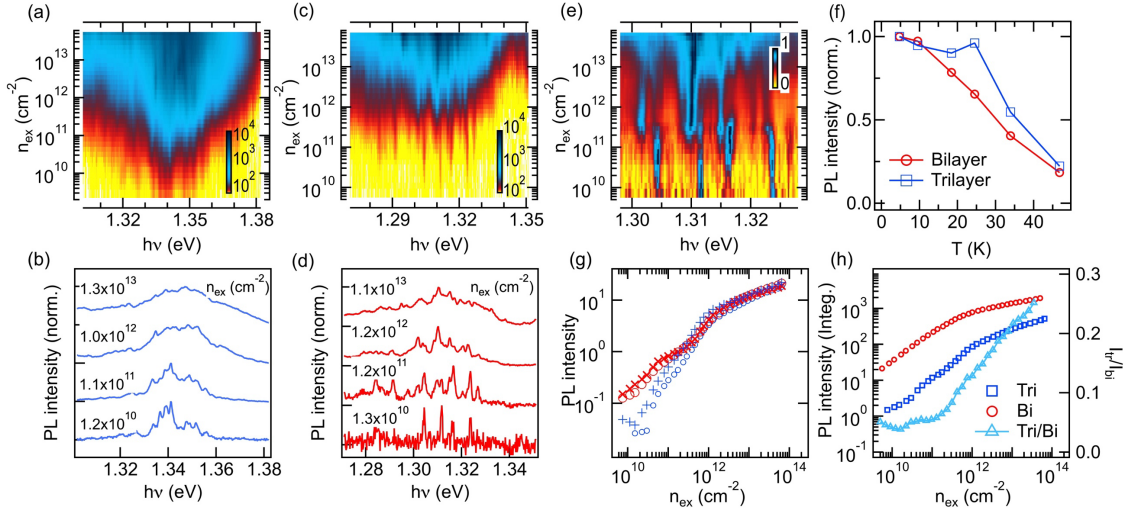

**Fig. 2 | Density-dependent PL spectra suggest interlayer exciton crystals in the trilayer. (a)** PL spectra as a function of $n_{ex}$ from interlayer excitons in the WSe$_2$/MoSe$_2$ bilayer. Color scale is PL intensity. **(b)** Comparative PL spectra at selected $n_{ex}$ from interlayer excitons in the WSe$_2$/MoSe$_2$ bilayer. **(c)** $n_{ex}$-dependent PL spectra from interlayer excitons in the WSe$_2$/MoSe$_2$/WSe$_2$ trilayer. **(d)** Comparative PL spectra at selected $n_{ex}$ from interlayer excitons in the WSe$_2$/MoSe$_2$/WSe$_2$ trilayer. **(e)** The same spectra as in **(c)**, but with PL intensity normalized to peak value at each $n_{ex}$ value. **(f)** Temperature-dependent PL intensity for interlayer excitons in the WSe$_2$/MoSe$_2$ bilayer (red circles) and the WSe$_2$/MoSe$_2$/WSe$_2$ trilayer (blue squares) at $n_{ex}$ = 6.3x10$^{11}$ cm$^{-2}$ and 7.9x10$^{11}$ cm$^{-2}$ at 4 K for the bilayer and the trilayer, respectively. **(g)** $n_{ex}$-dependent PL peak intensity for one of the peaks in **(e)** at 1.301eV (blue) / 1.304 eV (red) and 1.321 eV (blue) /1.324 eV (red), respectively. **(h)** Integrated PL intensities as functions of $n_{ex}$ from the bilayer ($I_B$, red circles) and the trilayer ($I_T$, blue squares), respectively. Also shown in **(h)** is the ratio $I_T/I_B$ (right axis). CW excitation: hν = 2.33 eV, sample temperature T = 4.6 K, except in panel (f) where T = 4.6 - 46 K.

**Robust exciton crystals and phase transitions in the trilayer.** The $n_{ex}$-dependent PL spectra provide evidence for the predicted exciton crystals in the trilayer. As comparison, we first present PL spectra from the WSe$_2$/MoSe$_2$ bilayer in a pseudo-color plot in Fig. 2a and at selected $n_{ex}$ in Fig. 2b. At the lowest $n_{ex}$, the spectra feature a group of narrow peaks attributed to moiré trapped excitons[11,13,28,29]. Above ~1x10$^{11}$ cm$^{-2}$, the peaks broaden and merge due to transition to the free exciton gas[15]; further broadening occurs upon transition to plasmas near and above the Mott density ($\sim 3 \times 10^{12}$ cm$^{-2}$), in agreement with previous studies[11,13,15].

The $n_{ex}$-dependent PL spectra from the WSe$_2$/MoSe$_2$/WSe$_2$ trilayer, Fig. 2c & 2d, are distinct. The first distinction is the extreme robustness of the sharp exciton peaks persisting to $n_{ex}$ in the $10^{12}$ cm$^{-2}$ region. This is more obvious in the peak intensity normalized PL spectra in Fig 2e. The peak positions and FWHMs remain nearly constant to $n_{ex} \sim 5 \times 10^{12}$ cm$^{-2}$, which is approximately the Mott density. These results suggest that crystal formation is more effective in spatially localizing the interlayer excitons and in preventing inter-exciton scattering than the moiré traps do. Supporting the stability of interlayer excitons in crystalline phases, we compare in Fig. 2f the temperature (T) dependences in integrated PL intensities at an excitation power of 0.5 μW, corresponding to $n_{ex} = 6.3 \times 10^{11}$ cm$^{-2}$ and $7.9 \times 10^{11}$ cm$^{-2}$ at 4 K for the bilayer and the trilayer, respectively. At this excitation power, the bilayer is dominated by interlayer excitons in the free exciton gas phase, and the PL intensity exhibits a rapid drop with increasing temperature. In contrast, the PL intensity remains nearly constant in the trilayer until a critical temperature of $T_c = 25$ K above which decay occurs. $T_c$ is likely the thermal melting temperature of the exciton crystals. The FWHM of PL from the trilayer remain narrow until T $\sim$ 25 K (Fig. S11-S12), consistent with the crystalline phases prior to thermal melting.

The second distinctive feature is the evolution of PL intensity with $n_{ex}$. At $n_{ex} \leq 2 \times 10^{11}$ cm$^{-2}$, the PL intensity from the trilayer is one-order of magnitude lower than that from the bilayer; this is consistent with the dominance of the predicted quadrupolar exciton lattice with low oscillation strength, as discussed above (Fig. 1f). These excitons become brighter with increasing $n_{ex}$ above $2 \times 10^{11}$ cm$^{-2}$ (close to $n_{moiré}$), as is most evident in the peak-intensity normalized spectra in Fig. 2e where clear onsets for the brightening of a portion of the excitons are seen. Fig. 2g compare as functions of $n_{ex}$ the integrated PL intensities of two pairs of excitons at 1.301eV/1.304 eV and 1.321 eV/1.324 eV, respectively. In each case, the intensity of the darker exciton (blue) catches up with that of the brighter ones at $n_{ex} > 2 \times 10^{11}$. The switching behavior is also reflected in the $n_{ex}$ dependences of PL intensity ratio between the trilayer and the bilayer, $I_{tr}/I_{bi}$, which remains constant (= 0.065±0.005) for $n_{ex} \leq 2 \times 10^{11}$ cm$^{-2}$ but increases continuously to $\sim$0.25 at $n_{ex} \sim 3 \times 10^{13}$ cm$^{-2}$, Fig. 1h. This trend suggests the transition and continuous addition to the bright dipolar exciton crystal in the trilayer, i.e., it is compressible. General features of the $n_{ex}$-dependent PL spectra are reproduced in device-II (Fig. S13).

Supporting the above interpretation, we note that PL emission of the trilayer region is red-shifted from that of the bilayer region by ~ 30 meV, in excellent agreement with theoretical prediction of the energetic difference between quadrupolar and dipolar excitons[8]. Moreover, the formation of quadrupolar excitons requires the hybridization of the top and the bottom $WSe_2$ monolayers through the middle $MoSe_2$. This hybridization is evidenced in reflectance spectra, Fig. S14. In contrast to the sharp and single transition in the $MoSe_2$ monolayer, spectra of the $WSe_2$ exciton transitions from the trilayer are characterized by broadening, red-shift, and peak splitting, consistent with interlayer hybridization. Further confirming the hybridization effect, we fabricated a twisted $WSe_2/WSe_2$ bilayer ($\theta = 3.0\pm0.5°$) with h-BN encapsulation and confirmed that hybridization results in broadening, red-shift, and peak splitting to the $WSe_2$ exciton transitions, in agreement with a recent report[35].

**Immobile excitons in the crystalline phases**. To determine the diffusivity of the interlayer excitons, we carry out confocal PL imaging under diffraction-limited and continuous wave (CW) excitation conditions[15], where the generation, diffusion, and recombination of the excitons reach a steady state. Fig. 3a shows PL intensity images of interlayer excitons in the bilayer (upper) and trilayer (lower) at selected excitation powers (0.1-3 μW), corresponding to average excitation densities within the diffraction limited spots of $n_{ex} = 1.7\times10^{11} - 2.2\times10^{12}$ cm$^{-2}$ and $n_{ex} = 2.2\times10^{11} - 3.0\times10^{12}$ cm$^{-2}$ for the bilayer and trilayer, respectively. At the lowest $n_{ex}$, the Gaussian width ($\sigma_{ex}$) of the PL image in either the bilayer or the trilayer is close to that of the excitation spot $\sigma_0 \sim 0.5$ μm, suggesting negligible diffusivity of the interlayer excitons (see Supporting Information Fig. S15-S18 from device-I and Fig. S19 from device-II). The Gaussian widths broaden with $n_{ex}$, but the trends are different for the two cases. Fig. 3b shows changes in the spatial variance of the PL images, $\Delta\sigma^2 = \sigma_{ex}^2 - \sigma_0^2$, as a function of $n_{ex}$. For interlayer excitons in the bilayer (red circles), $\Delta\sigma^2$ increases monotonically with $n_{ex}$, indicating increased diffusivity as the system transitions from moiré trapped interlayer excitons to free exciton gas and to electron-hole plasmas[15]. In contrast, $\Delta\sigma^2$ remains negligible at $n_{ex} < 5\times10^{11}$ cm in the trilayer, consistent with the dominance of crystalline phases. The abrupt change in the diffusivity at $n_{ex} > 5\times10^{11}$ cm$^{-2}$ is accompanied by changes in the emission spectra, i.e., the brightening of a large portion of the sharp PL peaks. This is particularly obvious in the intensity-normalized PL spectra in Fig. 2e and density-dependent PL intensities of selected peaks in Fig. 2g. These results reveal that the quadrupolar exciton crystal is completely immobile, while transition to the dipolar exciton phases (crystals and droplets) regains some mobility.

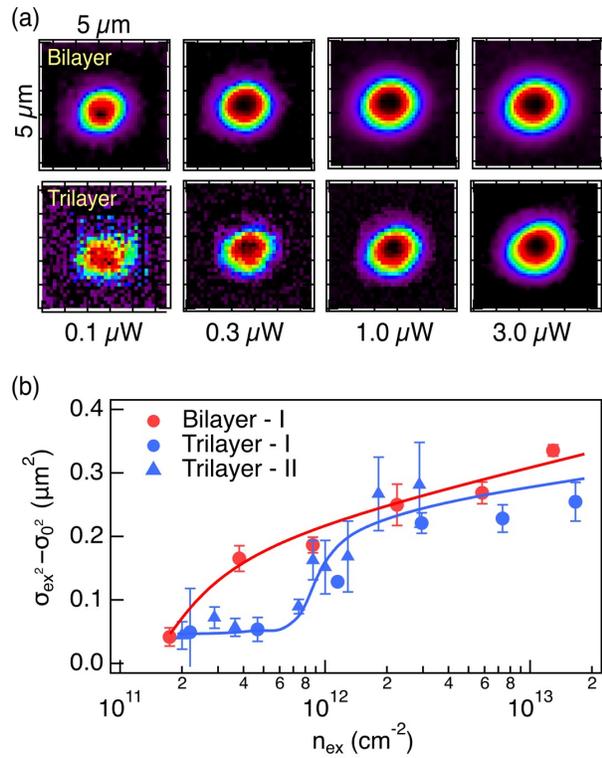

**Fig. 3 | Steady-state diffusion of interlayer excitons in $MoSe_2/WSe_2$ bilayer and $WSe_2/MoSe_2/WSe_2$ trilayer.** (**a**) Selected PL images at different CW excitation powers (0.1-3.0 μW) with a diffraction-limited excitation spot of Gaussian width $\sigma_0 \sim 0.5$ μm. Each image is 5x5 μm². (**b**) Excitation density ($n_{ex}$) dependent differential variance, $\Delta\sigma^2 = \sigma_{ex}^2 - \sigma_0^2$, where $\sigma_{ex}$ and $\sigma$ are the Gaussian widths of each PL image and the excitation spot, respective. Red dots: bilayer; blue dots: trilayer-I; blue triangles: trilayer-II. The red and blues curves serve as guides to the eye.

**Field induced excitonic phase transitions in the trilayer.** Further evidence of the exciton crystals comes from electric field ($\vec{E}$) dependences. Figs. 4a and 4b show $\vec{E}$-dependent PL spectra for the bilayer and the trilayer at $n_{ex} = 3.0\times10^{11}$ cm$^{-2}$ and $3.5\times10^{11}$ cm$^{-2}$, respectively. When $\vec{E}$ is applied in the perpendicular direction, interlayer exciton energy in the bilayer shifts linearly, Fig. 4a, as expected from a first-order Stark effect. Fitting the exciton peak shifts to $\Delta E_{ex} = -\vec{p}\cdot\vec{E}$ gives the interlayer exciton dipole moment of $\vec{p} = 5.7 \pm 0.1\ e\cdot$Å, in agreement with a previous report[33]. The $\vec{E}$-dependent PL spectra from the trilayer exhibit two linear regions with opposite slopes that symmetrically cross at $\vec{E} = 0$V, Fig. 4b, as expected from the Stark effect for dipolar excitons confined in the top or bottom heterojunctions, respectively, with oppositely pointing dipoles. From the slopes, we obtain $|\vec{p}| = 5.5 \pm 0.1\ e\cdot$Å in the trilayer. Note that the dipole moments of the interlayer excitons are identical for all the sharp PL peaks in either the bilayer or trilayer, as they are all determined from the interlayer separation at vdW contacts. In contrast, the PL peak positions vary due to sample disorder and changes in local electrostatic environments.

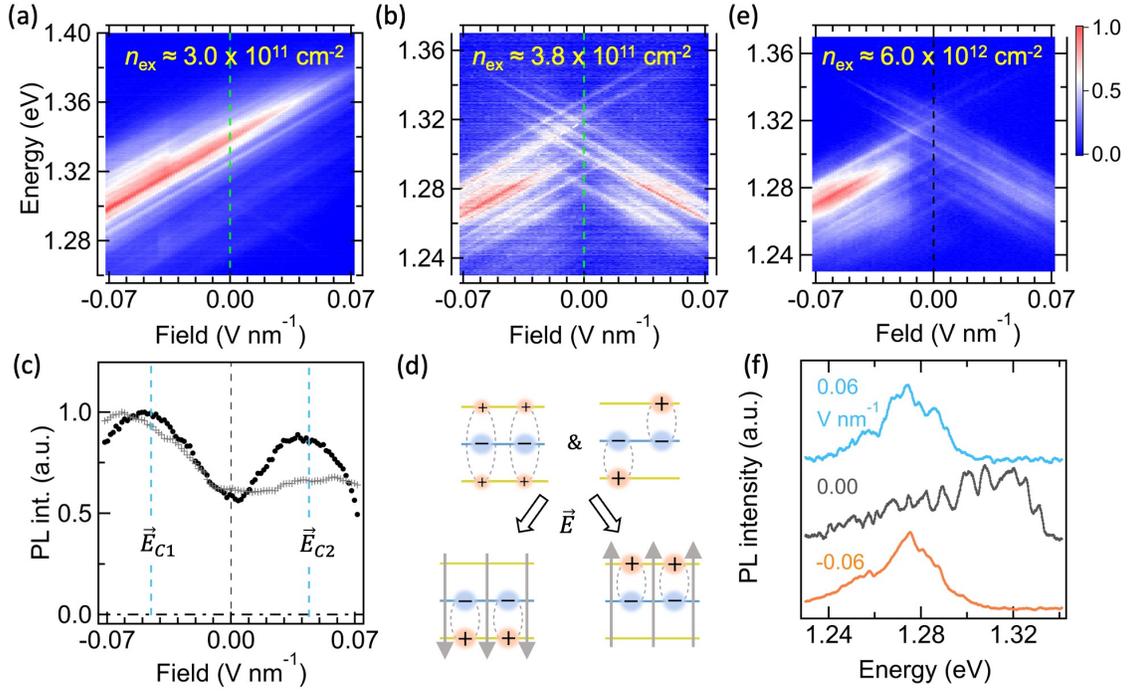

**Fig. 4 | Field-dependent PL spectra in the trilayer reveal dark-to-bright excitonic transition.** $\vec{E}$-dependent PL spectra of interlayer excitons in the WSe$_2$/MoSe$_2$ bilayer **(a)** and the WSe$_2$/MoSe$_2$/WSe$_2$ trilayer **(b, and e)**. The color bar represents normalized PL intensity. **(c)** Integrated PL intensity from the trilayer as a function of $\vec{E}$ in the trilayer. $\vec{E}_{C1}$ and $\vec{E}_{C2}$ label *(label $\vec{E}_{C1}$ and $\vec{E}_{C2}$ in panel c)* critical field values where the integrated PL intensities peak. The black dots are acquired by integrating the spectra in **(b)**, while the grey crosses are obtained by integrating the spectra in **(e)**. **(d)** Schematic illustration of quadrupolar to dipolar exciton transition induced by electric fields. **(f)** Comparative spectra of interlayer excitons in WSe$_2$/MoSe$_2$/WSe$_2$ acquired at electric field values of -0.06 V nm$^{-1}$ (orange), 0.00 V nm$^{-1}$ (gray), and 0.06 V nm$^{-1}$ (blue) sliced from the field-dependent data in **(e)**. This comparison highlights the disappearance of the fine spectral features due to the field-induced quantum phase transition from the staggered exciton lattice to electron-hole plasmas. CW excitation: hν = 2.33 eV; sample temperature T = 4.6 K.

Interestingly, the PL intensities (black dots in Fig. 4c) from the trilayer are lower at $|\vec{E}| = 0$, increase with $|\vec{E}|$, and reach peaks at $|\vec{E}| = 0.045\pm0.003$ V·nm$^{-1}$, before decreasing again at $|\vec{E}| > |\vec{E}_C|$. The same trends are observed for individual sharp peaks in the PL spectra (Fig. S20). A similar, albeit weaker, effect is reproduced in device-II (Fig. S21). Such a field-dependent change in PL intensity is absent in the bilayer region (Fig. S22-S24). The decrease in PL intensity with increasing $|\vec{E}|$ above $|\vec{E}_C|$ for red-shifted interlayer excitons can be understood from the decreased oscillator strength, which results from further spatial separation of the e-h pair driven by the $\vec{E}$ field[33]. The opposite trend seen in the region $|\vec{E}| < |\vec{E}_C|$ is unusual and suggests that excitons are are turned brighter by increasing $|\vec{E}|$. These observations are consistent with the presence of quadrupolar excitons, which are the most stable exciton phase with lower oscillator strength than that of dipolar excitons. This can be understood by the addition of the linear Stark effect, $-\vec{E}\sum_i \vec{p}_i$, to the system Hamiltonian in equation (2):

$$\hat{H}_{TL-E} = \sum_i (K_i) + \sum_{i<j}\left(V_{\uparrow_i \uparrow_j} - V_{\uparrow_i \downarrow_j}\right) - \Delta_Q \sum_i \sigma_i^x - \vec{E}\sum_i \vec{p}_i \qquad (3).$$

While the Stark term is zero for either quadrupolar or staggered dipolar lattices, it can overcome both stabilization energies ($V_{\uparrow_i \downarrow_j}$ and $\Delta_Q$) and switch the interlayer excitons to one of the two heterojunctions to align with $\vec{E}$, Fig. 4d. From the experimental values of $|\vec{E}_C|$ = 0.045 V·nm$^{-1}$, we have $\Delta E = -\vec{p} \cdot \vec{E}$ = -25 meV, which is within the range of the predicted stabilization energies of the exciton lattices[8,10].

At $n_{ex}$ > $n_{Mott}$ (~3x10$^{12}$ cm$^{-2}$ in the heterobilayer[15,18]), excitons in the trilayer driven by $\vec{E}$ to one of the two heterobilayers is expected to undergo a Mott transition. This is confirmed in Fig. 4e for the trilayer at $n_{ex}$ = 6.0x10$^{12}$ cm$^{-2}$, where the sharp PL peaks at $\vec{E}$ = 0 V·nm$^{-1}$ merge into broadened peaks as $\vec{E}$ is swept in both directions, indicating field-driven phase transitions to electron/hole plasmas. Fig. 4f compares the PL spectra at $\vec{E}$ = 0, 0.6, and -0.6V. The sharp PL peaks of the crystalline phases at $\vec{E}$ = 0 V broaden into spectra characteristic of the plasma phase in one of the two bilayers at $\vec{E}$ = 0.6 or -0.6V. For comparison, in the bilayer region at $n_{ex}$ > $n_{Mott}$, the PL spectra remain broad as expected from the plasma phase in the entire range of Stark tuning; see Fig. S22 for $n_{ex}$ = 4.8x10$^{12}$ cm$^{-2}$. Note the critical field $|\vec{E}|$ for the transition from the quadrupolar to the brighter dipolar excitons shifts to higher values at larger $n_{ex}$ (grey dots in Fig. 4c for $n_{ex}$ = 6.0×10$^{12}$ cm$^{-2}$); this may be attributed to the higher cohesive energy of the quadrupolar exciton crystals at higher $n_{ex}$. In related work, a recent posting by Li et al. also presented evidence for the conversion of quadrupolar to dipolar interlayer excitons by vertical electric fields in the system of WS$_2$/WSe$_2$/WS$_2$ heterotrilayer, consistent with ab initio calculations [36].

Note that the asymmetric field-dependent spectral intensity in Fig. 4e likely originates from device asymmetry. The top graphite electrode may be closer to the heterotrilayer than the bottom electrode is; this would result in more efficient quenching of dipolar excitons located in the top heterobilayer due to Forster energy transfer. Moreover, sample heterogeneity and closeness to the spatially connected trilayer region can contribute to the variation in field-dependent PL intensities from the bilayers (Fig. 4a and Fig. S22-S23).

**Concluding remarks.** We present experimental evidences for ordered phases of interlayer excitons, as theoretically predicted for the symmetric $WSe_2/MoSe_2/WSe_2$ heterotrilayer. Ordered quantum phases in TMDC bilayers have been reported before for Fermions, and our experimental evidences for the bosonic interlayer excitons in the trilayer likely result from the increased "dimensionality" which stabilizes the ordered bosons. Similar stabilization has also been reported most recently for Bose-Fermi mixtures of interlayer excitons and holes in the $WS_2$/bilayer-$WSe_2/WS_2$ heterostructure [37]. An intriguing question for future research is whether ordering of excitonic quasiparticles can lead to quantum coherent phenomena[38], such as exciton condensation[39,40] or Dicke superradiance in ordered emitter arrays[41,42]. The study of these coherent phenomena requires the determination of the quantum nature of light from correlation analysis, which is difficult due to the low oscillator strength of interlayer excitons in the $WSe_2/MoSe_2$ heterojunction. In this regard, it might be beneficial to use other symmetric trilayers, such as $WS_2/MoSe_2/WS_2$, where hybridization of interlayer exciton with intralayer exciton gives rise to much larger oscillator strength[14]. These and other TMDC multilayers may provide versatile playgrounds for the exploration of quantum phase transitions and quantum coherent phenomena of optically bright bosons.

**Methods**

**Device fabrications.** Monolayer WSe2 and MoSe2 were mechanically exfoliated from bulk crystals grown by the self-flux method. These monolayers feature the lowest reported defect densities ($\leq 10^{10}$ cm$^{-2}$)[43]. Flakes of h-BN with thicknesses of 11-13 nm and graphite with thicknesses of ~1–1.5 nm (~ 3-4 layers of graphene, optically transparent) were also obtained via mechanical exfoliations. The surfaces flatness (height fluctuation < 0.5 nm across the whole flake) of TMDC monolayer, h-BN flakes, and thin graphite were all characterized by atomic force microscopy (AFM).

The h-BN/Graphite (top-gate)/h-BN/$WSe_2/MoSe_2/WSe_2$/Graphite (contact)/h-BN/Graphite (bottom-gate)/h-BN heterostructure was prepare by the dry stamp method[44]. Briefly, we coated a thin polycarbonate (PC) polymer film on top of a transparent polydimethylsiloxane (PDMS) cube (~1×1×1 mm³), and used this micro-lens stamp to pick up the top h-BN flake. This h-BN was then used to pick up all the rest of the 2D flakes and monolayers one-by-one in a desired sequence. The near 0° twist angles between WSe$_2$ and MoSe$_2$ monolayers were ensured by SHG characterization (*vide infra*), and aligned via a rotation stage. A key feature of the trilayer device is structural symmetry. In order to meet this requirement, we employ the tear-and-stack approach to fabricate the heterostructure, i.e., one single sheet of WSe$_2$ was utilized; this WSe$_2$ monolayer was first teared and one of the two teared piece was picked up, followed by picking up the MoSe$_2$ layer, and then the second teared piech of WSe$_2$ to form the WSe$_2$/MoSe$_2$/WSe$_2$ heterostructure (see Fig. S4 for an illustration). After picking up all the layers in the desired order, we transferred the entire heterostructure onto a clean silicon wafer at elevated temperatures (~120 – 180°C). The sample surface was then washed with chloroform, acetone, and isopropanol consecutively. The heterostructure was then patterned with metal electrode: metal gates and contacts (Cr/Pd/Au: 3nm/20nm/60nm) were evaporated (Angstrom Engineering UHV E-beam deposition system) on top of the exposed regions of the gate and contact graphite (see SI Fig. S5 for the full optical image of the device).

**Determination of the TMDC monolayer crystal orientation by SHG.** The detailed procedures for polarization-resolved SHG measurements were discussed in ref. [45]. We acquired the azimuthal angular (θ) distribution of SHG signal by rotating the laser polarization and the SHG signal (via a half waveplate) at fixed sample orientation. Due to the D$_{3h}$ symmetry, the non-vanishing tensor elements of the second order susceptibility of WSe$_2$ or MoSe$_2$ monolayer are $\chi_{yyy}^{(2)} = -\chi_{yxx}^{(2)} = -\chi_{xxy}^{(2)} = -\chi_{xyx}^{(2)}$ where the *x* axis is defined as the zigzag direction.[45] When we simultaneously rotated the fundamental and SHG signals, the SHG intensity showed six-fold symmetry: $I_\perp \propto cos^2(3\theta)$ and $I_\parallel \propto sin^2(3\theta)$, where $\theta$ is the angle between the laser polarization and the zigzag direction. We fixed all substrates supporting the TMDC monolayers are onto a single glass slide; thus, the relative crystal orientations are determined directly by comparison to a triangular CVD TMDC monolayer where the zigzag directions are their crystal edges.

To perform the phase-resolved SHG measurements, Fig. S3, we employed a similar strategy as that described elsewhere[46,47]. Prior to phase-resolved measurements, we assured that the azimuthal angles of TMDC monolayers were co-polarized (relying on the above polarization-resolved SHG measurements); however, at this stage, it remained unknown whether these co-polarized TMDC monolayers possessed 0° or 60° twist angles. Linearly polarized femtosecond laser light (Spectrum Physics Tsunami, 80 MHz, 800 nm, 80 fs) was first focused onto a thin z-cut quartz crystal (0.1 mm × 1 cm × 1 cm, MTI Corporation) by a NIR AR-Coated lens (f = 5 cm), which generated a 400 nm SHG reference pulse signal (SHG$_1$) propagating with the remaining 800 nm femtosecond pump beam. These beams were collected by an achromatic (400-1100 nm) doublet lens and collimated. A picosecond-scale time interval ($\Delta\tau$) between the SHG$_1$ and pump pulse existed. The beams were directed to an inverted optical microscope (Olympus IX73) where the samples were placed. Upon the arrival and focus (NA 0.50 objective, Olympus UPLFLN 20X) of the two pulses at the TMDC monolayer sample surface, another SHG signal (SHG$_2$) was produced from the TMDC monolayer by the remaining 800 nm pump beam. SHG$_1$ and SHG$_2$ were collectively filtered by a short-pass filter and directed to a visible spectrometer, and further detected by an EM-CCD camera. The diffraction grating performed the time-to-frequency domain Fourier transformations for SHG$_1$ and SHG$_2$ with interval $\Delta\tau$, which was followed by the spectral interferences at the EM-CCD; the ultimate spectra from the EM-CCD showed the interference patterns. In this regard, TMDC monolayers having a $\theta$ initial azimuthal angle vs. that having a $\theta$+60° azimuthal angle displayed alternating interference patterns as shown in SI, Fig. S2.

**Electric field and doping**. Electric field application and electrostatic doping in a parallel dual-gate device were discussed in detail previously[48]. The equivalent circuit of our dual-gate device was a parallel plate capacitor with a 2D sheet of material (i.e., the TMDC heterostructure) in the middle. We define the thickness of the top insulating h-BN in between the TMDC heterostructure and the top graphite gate as $d_1$ ($\approx$ 11 nm), and the bottom one as $d_2$ ($\approx$ 13 nm). The dielectric constants for h-BN and TMDC are $\varepsilon_{h-BN} \approx 3.9$, and $\varepsilon_{TMDC} \approx 7.2$.[48–50] The electric field generated in between the top and bottom gates can be expressed as

$$|\vec{E}| = \frac{V_{tg}-V_{bg}}{d_1+d_2}/(\frac{\varepsilon_{TMDC}}{\varepsilon_{h-BN}} + \frac{d_{HS}}{d_1+d_2}) \tag{4}$$

where $V_{tg}$ and $V_{bg}$ are the top and bottom gate voltages, respectively; $d_{HS}$ is the thickness of the TMDC heterostructure. The doping in the middle 2D sheet material (i.e., TMDC trilayer) is

$$ne \approx C_1 V_{tg} + C_2 V_{bg} - n_0 e \approx \frac{\varepsilon_{h-BN}\cdot A}{d_1}V_{tg} + \frac{\varepsilon_{h-BN}\cdot A}{d_2}V_{bg} - n_0 e \tag{5}$$

where $n$ is the doping density, $C_1$ ($C_2$) is the capacitance from the capacitor formed by the top (bottom) graphite gate and the TMDC sheet, $A$ is the capacitor area, and $n_0$ is the in-gap state density that need to be filled before the conduction (valence) band is filled with electrons (holes). We control the electric field and electrostatic doping independently: (1) if $V_{tg} = -\frac{d_1}{d_2}V_{bg}$, there will be no external doping effect, and $|\vec{E}|$ scales linearly with $V_{tg} - V_{bg}$; (2) if $V_{tg} = V_{bg}$, there will be no external electric field, and doping density scales linearly with $V_g$ ($= V_{tg} = V_{bg}$). Additionally, we treat the energy shift ($\Delta E$) of dipolar excitons in an electric field based on the first-order Stark effect,

$$\Delta E = -\vec{p} \cdot \vec{E} \qquad (6)$$

where $\vec{p} = e \cdot \vec{d}$, $\vec{d}$ is the interlayer separation between two TMDC monolayers with the direction pointing from the negative to positive charge. This enabled us to estimate the electron-hole separation to be ~ 6Å based on the $\vec{E}$-dependent PL spectra.

**Photoluminescence microscopy/spectroscopy.** The gate-dependent PL spectra measurements were performed on a home-built spectro-microscopic system[45] based on a liquid-helium recirculating optical cryostat (Montana Instruments Fusion/X-Plane) with a 100x, NA 0.75 objective (Zeiss LD EC Epiplan-Neofluar 100x/0.75 HD DIC M27). The temperature of the sample stage could be varied between 3.8 K and 350 K. In all experiments presented in this study, the TMDC heterobilayer and monolayers samples were at 4 K in a vacuum (<10$^{-6}$ torr) environment, unless otherwise noted. The incident laser beam (CW, 532 nm in all experiments, except TCSPC) was focused by the objective to a diffraction limited spot on the sample. The excitation power was measured by a calibrated power meter (Ophir StarLite) with broad dynamic range. The PL light was collected by the same objective, spatially and spectrally filtered, dispersed by a grating, and detected by an InGaAs photodiode array (Princeton Instruments PyLoN-IR). The wavelength was calibrated by neon-argon and mercury atomic emission sources (Princeton Instruments IntelliCal). In all PL measurements, we found no laser heating of the sample under cryogenic cooling. The PL spectra are completely reproducible following repeated measurements at the same spot on the sample. The sample was grounded during the measurements. Gate voltages were supplied via two identical source meters (Keithley 2400, source accuracy ~ 0.02%) which were computer controlled through a GPIB bus.

In time-resolved photoluminescence (TRPL) measurements, the pulsed excitation light (hν = (1.65 eV, pulse duration 150 fs) was from a wavelength tunable output of an visible optical parametric amplifier (Coherent OPA 9450) pumped by a Ti:Sapphire regenerative amplifier (Coherent RegA 9050, 250 kHz, 800 nm, 100 fs). The interlayer PL emission was focused onto a single photon avalanche photodiode (IDQ ID100-50). The TRPL trace was collected with a time-correlated single photon counting (TCSPC) module (SPC-130). The instrument response function, determined by collecting scattered laser light, has a full-width-at-half-maximum (FWHM) of 100 ps. The time resolution of TRPL was estimated at ~20% of the FWHM, i.e. ~20 ps.

**Acknowledgements**. This work was supported by the Materials Science and Engineering Research Center (MRSEC) through NSF grant DMR-2011738. Partial supports for the fabrication and characterization of device-I by Programmable Quantum Materials, an Energy Frontier Research Center funded by the U.S. Department of Energy (DOE), Office of Science, Basic Energy Sciences (BES), under award DE-SC0019443 and for device-II and device-III by the Vannevar Bush Faculty Fellowship through the Office of Naval Research through Grant No. N00014-18-1-2080 are also acknowledged. We thank Andrew Millis, Ana Asenjo-Garcia, Xiaodong Xu, and Farhan Rana for fruitful discussions and Wenjing Wu for help with sample preparation at the initial stage of this project.

**Data Availability**. The data represented in Figs. 1-4 are provided with the article source data. All data that support the results in this article are available from the corresponding author upon reasonable request.

**Author contributions:**

YB and XYZ conceived this work. YB and YL carried out all experiments with assistance by SL, YG, JP, JW, CRD, and JH. XYZ supervised the project. The manuscript was prepared by YJB and XYZ in consultation with all other authors. All authors read and commented on the manuscript.

**Competing Interests.** All authors declare that they have no competing interests.